\documentclass[%
 reprint,
 amsmath,amssymb,
 aps,
 prl,
]{ revtex4-2}

\usepackage{graphicx}% Include figure files
\usepackage{dcolumn}% Align table columns on decimal point
\usepackage{bm}% bold math
\usepackage{epstopdf}
\usepackage{color}
\usepackage{ulem} 
\begin{document}

\preprint{APS/123-QED}

\title{Experimental Realization of All-Optical Terahertz Attoclock}% Force line breaks with \\

\author{Yanjun Gao$^{1,2,5}$}
\author{Yizhu Zhang$^{3}$}
\email{zhangyizhu@tiangong.edu.cn}
\author{Yulin Chen$^{4}$}
\author{Jingjing Zhao$^{4}$}
\author{Meng Li$^{2}$}
\author{Xiaokun Liu$^{2}$}
\author{Yange Chen$^{1,5}$}
\author{Jiahui Guo$^{2}$}
\author{Kiyoshi Ueda$^{2,6}$}
\author{Ahai Chen$^{2}$}
\email{chenah1@shanghaitech.edu.cn}
\author{Yuhai Jiang$^{2,4,1}$}
\email{jiangyh3@shanghaitech.edu.cn}

\affiliation{
$^1$Shanghai Advanced Research Institute, Chinese Academy of Sciences, Shanghai 201210, China\\
$^2$Center for Transformative Science and School of Physical Science and Technology, ShanghaiTech University, Shanghai 201210, China\\
$^3$School of Electrical Engineering, Tiangong University, Tianjin 300387, China\\
$^4$School of Physics, Henan Normal University, Xinxiang 453007, China\\
$^5$University of Chinese Academy of Sciences, Beijing 100049, China \\
$^6$Department of Chemistry, Tohoku University, No. 41 Kawauchi, Sendai, 980-8578, Japan
}

% \affil[7]{\orgdiv{Department of Chemistry}, \orgname{Tohoku University}, \orgaddress{\street{No. 41 Kawauchi}, \city{Sendai}, \postcode{980-8578}, \country{Japan}}}

% \collaboration{CLEO Collaboration}%\noaffiliation

\date{\today}% It is always \today, today,

\begin{abstract}

The attoclock is a powerful tool for probing ultrafast electron dynamics with attosecond precision. Here, we demonstrate an all-optical terahertz (THz) attoclock that reconstructs  photoionization dynamics by detecting the THz radiation emitted from Ar atoms ionized by
%the fundamental and second harmonic 
two-color (800 nm/400 nm) laser fields.
%, without the need for photoelectron detection. 
%
In this approach, the polarization direction of the emitted THz field reflects the direction of the photoelectron drift velocity and thus serves as a direct observable that encodes the effective ionization delay, analogous to the angular deflection of photoelectrons in conventional attoclocks. 
By precisely tailoring the relative phase and ellipticity of the driving fields, we observe intensity-dependent rotations of the THz polarization.
These rotations, which reveal changes of the effective delay, are consistent with 
both conventional attoclock measurements and time-dependent Schr\"odinger equation simulations.
Our experiment establishes the feasibility of the THz attoclock as a vacuum-free and contactless probe of tunneling dynamics, offering a transformative alternative for investigating condensed-matter systems where photoelectron detection is challenging.

%with promising applications for condensed matter systems where photoelectron detection is impractical.

\end{abstract}

\maketitle

%\tableofcontents

%\section{}

The photoelectron attoclock technique employs a single-color laser pulse~\cite{1attoclock2008_np,1attoclock2008_science,1attoclock2012_KellerUrsula,1attoclock2017_Robert,1attoclock_H_2019,1attoclock2024_coulomb_DingDajun,1attoclock2024_liuxiaojun,1attoclock2025_LiWen} or a two-color laser pulse~\cite{2attoclock2019_Liuyunquan_theory,2attoclock2021_Liuyunquan_POP,2attoclock2024_Lupeixiang} to map the ionization time of an electron to its emission angle in the photoelectron momentum distribution (PMD). This technique has been widely used to probe photoelectron ionization dynamics, such as the tunneling time and correlated-electron emission time,
\iffalse
It has been applied to investigate tunneling time~\cite{1attoclock2008_np,1attoclock2008_science,1attoclock2017_Robert,1attoclock2017_Robert,1attoclock_H_2019,2attoclock2021_Liuyunquan_POP}, 
%~\cite{attoclock2008_science,Olga2015_attoclock_theory_instantaneous,attoclock2014_Ursula_Optica,attoclock5,attoclock_H_2019,attoclock2019_Ursula_revisit,attoclock8,Liu2019_attoclock_theory}, 
tunneling geometry~\cite{1attoclock2012_KellerUrsula}, and Coulomb effect~\cite{1attoclock2024_coulomb_DingDajun}, thereby advancing our understanding of the strong-field ionization dynamics.
\fi
%
driving advances in attosecond metrology and providing new insights into laser-driven electron dynamics in atoms and molecules.

\begin{figure}[tb]
\centering
\includegraphics[width=8.5cm,height=5.36cm]{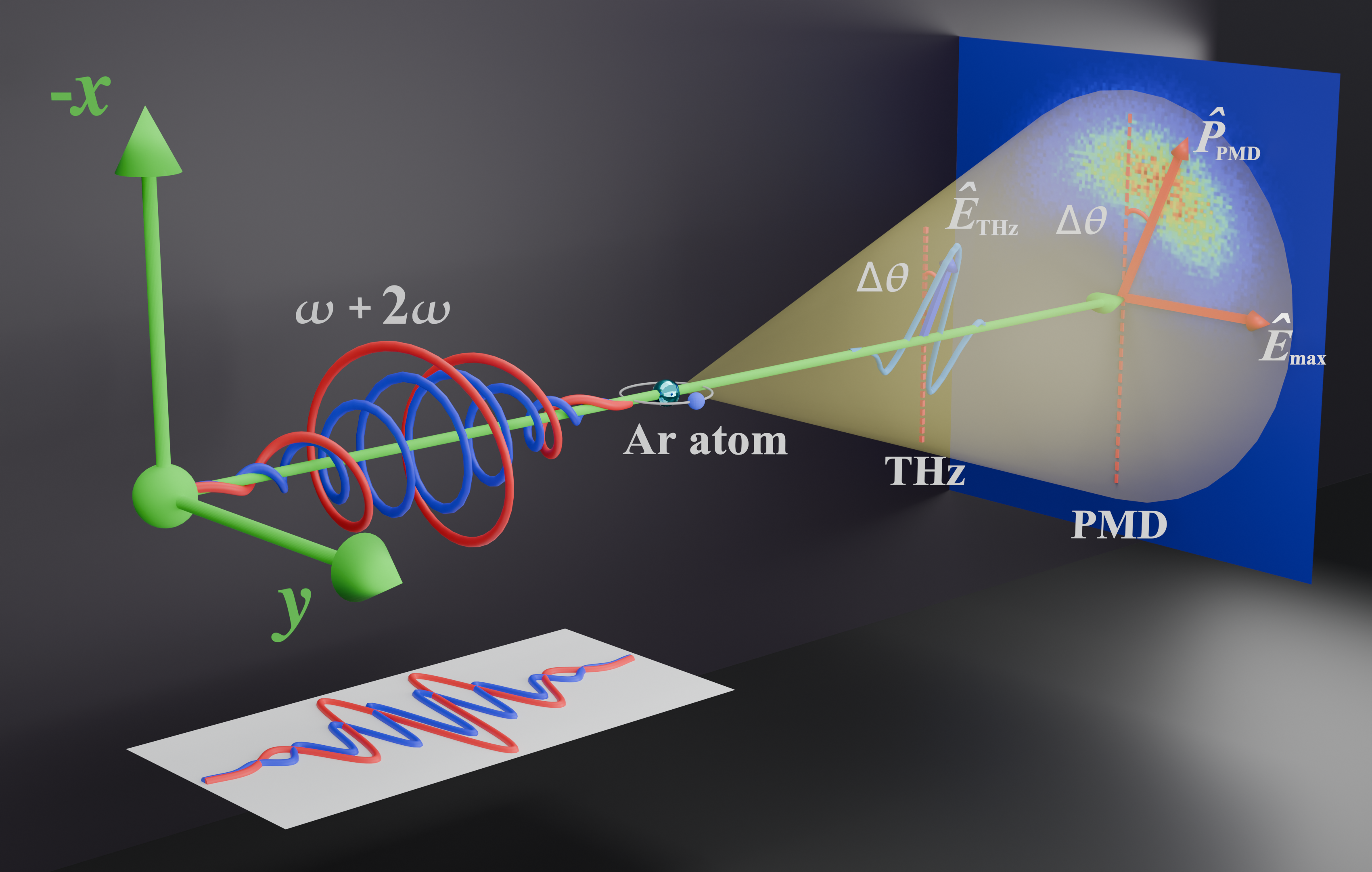}
\caption{\label{fig1} Schematic of THz attoclock. A fundamental field with an ellipticity of $\epsilon_\omega\approx 0.81$ and its major axis direction along the $y$ axis, together with a circularly polarized second harmonic field ($\varphi=0$), jointly ionize Ar atoms. 
%The relative phase of the two-color fields is zero. 
The ionized photoelectrons are accelerated and radiate THz waves under the action of the two-color fields. 
%The THz attoclock is equivalent to the photoelectron attoclock based on the PMD. 
Because the drift velocity of tunneling electrons accelerated in the two-color fields determines the magnitude and direction of the radiated THz field, the THz polarization direction ($\hat{\boldsymbol{E}}_{\mathrm{THz}}$) aligns with the asymmetric direction of the PMD ($\hat{\boldsymbol{P}}_{\mathrm{PMD}}$). 
%Equivalent to photoelectron attoclock, the time when electrons appear in the continuous state can also be mapped to the deflection of $\hat{\boldsymbol{E}}_{\mathrm{THz}}$. 
In analogy to the photoelectron attoclock, the time when electrons appear in the continuum can also be mapped to the deflection of $\hat{\boldsymbol{E}}_{\mathrm{THz}}$.
The angular offset $\Delta\theta$ refers to the angle between $\hat{\boldsymbol{E}}_{\mathrm{THz}}$ (or $\hat{\boldsymbol{P}}_{\mathrm{PMD}}$) and the minor axis of the $\omega$ field, which can be mapped to the effective delay. $\hat{\boldsymbol{E}}_{\mathrm{max}}$ refers to the direction at the peak of the total electric field. 
%251101，相对相位为0时，峰值总电场沿基频场长轴方向，此时才可以说∆θ是指"THz偏振方向"和"短轴"的夹角。所以我认为还是要说明相对相位为0。-------gyj  
%这里有点啰嗦了，能不能直接写$\varphi =0 $
%251101，老师，"equivalent delay"译为等效延迟，NP文中为"effective  delay".咱们也统一改为"effective  delay"吧？----gyj
%可以
%251101，目前这幅图里的PMD是CTMC算的结果，没有干涉现象。之后TDSE的PMD出来的话，会换上TDSE的结果。---gyj
%不用换，这个是示意图。不要把时间浪费在这里。
}
\end{figure}

The direct ionization of photoelectrons is inherently accompanied by the generation of terahertz (THz) radiation, originating from continuum–continuum transition of the released electrons~\cite{zhangPR,zhangOL}. This emission process establishes an intrinsic connection between the microscopic electronic dynamics and the macroscopic THz field, which encodes information about the photoelectron trajectory and the ionization timing~\cite{ganPRA,ganOE}. 
Recently, this connection has been theoretically reaffirmed~\cite{alloptical,ganPRA}, leading to the proposal of an all-optical attoclock in the THz regime (THz attoclock).
%
%Recent studies have theoretically proposed a scheme for an optical attoclock in the terahertz regime (THz attoclock)~\cite{alloptical,ganPRA}. 
The THz attoclock circumvents the need for photoelectron detection by measuring the THz radiation induced by the acceleration of the photoelectron.
%There have been previous reports using the high-harmonic radiation to study ionization time~\cite{HHG1,HHG2,HHG3,HHG4}. The challenge with this method is the need to decouple the ionization and recombination steps~\cite{alloptical}. 
%高次谐波我觉得就别说了，影响整篇论文的逻辑。
%下面不要用ctmc的公式，会让读者认为以下的规律只有在特定模型下有效。
The schematic of THz attoclock is illustrated in Fig.~\ref{fig1}. A fundamental-frequency ($\omega$) laser field (red waveform) and its second harmonic (2$\omega$, blue waveform) ionize Ar atoms, and the resulting photoelectrons are subsequently accelerated in the combined $\omega$-$2\omega$ fields, leading to dipole radiation.
%251110，老师，我把文中所有的"w和-"之间都加了空格---gyj
Because the $\omega$-$2\omega$ fields are nearly circularly polarized, the electrons are driven far away from the parent ion. Consequently, radiation does not come from high-harmonic generation through electron–ion recombination~\cite{HHG2,zhang2012PRL_zzx_yjm}.
Instead, it originates from the transient acceleration of photoelectrons—known as Brunel radiation~\cite{bruneharmonic1,bruneharmonic2}—which contributes to low-order harmonic generation. The lowest-frequency component of Brunel radiation corresponds to the THz emission (light-blue waveform in Fig.~\ref{fig1}).

The connection between THz emission and PMD is illustrated in Fig.~\ref{fig1}. The PMD represents the probability $\mathcal{P}$ of photoelectrons with final velocity $\boldsymbol{v}(t\rightarrow\infty)$. Electrons accelerated in asymmetric $\omega$-$2\omega$ fields produce an asymmetric PMD, whose asymmetry is defined as~\cite{ganPRA}
\iffalse
\begin{align}
    \boldsymbol{P}_{\mathrm{PMD}} = \boldsymbol{v}_{\mathrm{int}}(t=\infty)= \int_{\boldsymbol{v}}  w(\boldsymbol{v}(t = \infty)) \boldsymbol{v}(t = \infty) d \boldsymbol{v}, \nonumber
\end{align}

%251110，老师，一份"动量乘以权重"为"w(v)*dv",所有"w(v)*dv"相加，被积函数应该是w(v)-------gyj
%
{\color{blue}@Ahai  use "probability" instead of "weight" sounds more general here
\begin{align}
    \boldsymbol{P}_{\mathrm{PMD}} = \boldsymbol{v}_{\mathrm{int}}(t\rightarrow\infty)= \int_{\boldsymbol{v}}  \mathcal{P}[\boldsymbol{v}(t \rightarrow \infty)] \hat{\boldsymbol{v}}(t \rightarrow \infty) d \boldsymbol{v}, \nonumber
\end{align}
}
%
\fi
\begin{align}
    \boldsymbol{P}_{\mathrm{PMD}} = \boldsymbol{v}_{\mathrm{int}}(t\rightarrow\infty)= \int_{\boldsymbol{v}}  \mathcal{P}[\boldsymbol{v}(t \rightarrow \infty)] d \boldsymbol{v}, \nonumber
\end{align}
where $\boldsymbol{v}_{\mathrm{int}}$ denotes the integrated velocity of the emitted photoelectrons.
The asymmetric direction of PMD $\hat{\boldsymbol{P}}_{\mathrm{PMD}}$ is indicated by the red arrow in Fig.~\ref{fig1}. The Brunel harmonic radiation originates from the transient acceleration of photoelectrons, expressed as $  \boldsymbol{E}_{\mathrm{Brunel}}(t) \propto \frac{\partial \boldsymbol{v}_{\mathrm{int}}(t)}{\partial t}$.
Its lowest-frequency component corresponds to the THz emission and can be expressed as~\cite{alloptical,ganPRA} 
\begin{align}
    \boldsymbol{E}_{\mathrm{THz}}  
    &\propto \lim_{\nu_\text{} \to 0}
\int_{-\infty}^{\infty}
\frac{\partial \boldsymbol{v}_{\mathrm{int}}(t)}{\partial t}
e^{-i2\pi \nu_\text{}t} dt 
=\boldsymbol{v}_{\mathrm{int}}(t\rightarrow\infty). \nonumber 
% &= \boldsymbol{v}_{\mathrm{int}}(t = \infty). \nonumber
\end{align}
By controlling the relative phase $\varphi$ of the fields $\omega$-$2\omega$, the resulting asymmetric fields induce a non-zero $\boldsymbol{v}_{\mathrm{int}}(t\rightarrow\infty)$, i.e., a residual photocurrent~\cite{kimPC1,kimPC2}%这里请引用光电流的文献
, which gives rise to the THz radiation. 
%$\varphi$ is the phase of the $2\omega$ field relative to the $\omega$ field.
%
%Evidently, the polarization direction of the THz field, $\hat{\boldsymbol{E}}_{\mathrm{THz}}$ (blue arrow in Fig.~\ref{fig1}), coinciding with $\hat{\boldsymbol{P}}_{\mathrm{PMD}}$, is consistent with the direction of $\boldsymbol{v}_{\mathrm{int}}(t\rightarrow\infty)$. 
%
%{\color{blue} 
Evidently, the polarization direction of the THz field $\hat{\boldsymbol{E}}_{\mathrm{THz}}$ (blue arrow in Fig.~\ref{fig1}) coincides with $\hat{\boldsymbol{P}}_{\mathrm{PMD}}$.
%{\color{blue} Evidently, the polarization direction of the THz field $\hat{\boldsymbol{E}}_{\mathrm{THz}}$ (blue arrow in Fig.~\ref{fig1}) $\hat{\boldsymbol{P}}_{\mathrm{PMD}}$ are both consistent with the direction of $\boldsymbol{v}_{\mathrm{int}}(t\rightarrow\infty)$, and therefore with each other.
%{\color{blue} Evidently, both consistent with the direction of $\boldsymbol{v}_{\mathrm{int}}(t\rightarrow\infty)$, the polarization direction of the THz field, $\hat{\boldsymbol{E}}_{\mathrm{THz}}$ (blue arrow in Fig.~\ref{fig1}), coincides with $\hat{\boldsymbol{P}}_{\mathrm{PMD}}$.
%}
%
%In a conventional photoelectron attoclock, the tunneling time delay is encoded in the angular deflection of $\hat{\boldsymbol{P}}_{\mathrm{PMD}}$.
%
In a conventional photoelectron attoclock, 
%the angular deflection of $\hat{\boldsymbol{P}}_{\mathrm{PMD}}$ integrates contributions from tunneling ionization delay, non-adiabatic effects, and long-range Coulomb interactions, all of which are encapsulated into an effective delay $\tau$~\cite{alloptical}.
the angular deflection of $\hat{\boldsymbol{P}}_{\mathrm{PMD}}$ may arise from the contributions of tunneling ionization delay, non-adiabatic effects, and long-range Coulomb interactions, which can be converted into an effective delay $\tau$~\cite{alloptical}.
%
%
%Since $\hat{\boldsymbol{E}}_{\mathrm{THz}}$ is equivalent to $\hat{\boldsymbol{P}}_{\mathrm{PMD}}$, $\hat{\boldsymbol{E}}_{\mathrm{THz}}$ can therefore serve as an all-optical proxy for reconstructing the tunneling delay of photoelectrons.
%{\color{blue}
Therefore, $\hat{\boldsymbol{E}}_{\mathrm{THz}}$ can serve as an all-optical proxy for reconstructing the effective delay of photoelectrons. Note that $\tau$ is treated here as a cumulative timing observable, without making a distinction among the various theoretical definitions of tunneling time.
While a nearly circularly polarized two-color driving field has been theoretically proposed to implement a THz attoclock~\cite{ganPRA}, its experimental realization remains challenging. A primary technical hurdle involves the precise control and calibration of the relative phase $\varphi$ between the $\omega$ and 2$\omega$ fields across varying laser intensities, which is essential for accurately determining the subtle angular deflection of $\hat{\boldsymbol{E}}_{\mathrm{THz}}$. In addition, laser ellipticity must be finely tuned to avoid electron–ion recollision and ensure a clear interpretation of the THz attoclock~\cite{ganPRA}.
%
%Furthermore, THz generation in the $\omega$-$2\omega$ scheme involves multiple coupled processes of photoionization, nonlinear self-focusing and defocusing, and THz absorption losses in plasma~\cite{effect2,effect5,twofoci}, which may complicate the THz attoclock measurements.
Beyond these technicalities, THz generation in the $\omega$-$2\omega$ scheme is inherently coupled with complex nonlinear phenomena, including self-focusing, plasma defocusing, and absorption losses~\cite{effect2,effect5,twofoci}. These effects may distort the driving-field waveform and shift the relative phase $\varphi$ at the focus, leading to a deformation of the $\varphi$-dependent THz electric-field curves, thereby complicating THz attoclock measurements. Whether the subtle THz polarization deflection induced by intrinsic atomic ionization dynamics can survive these intertwined propagation effects is a pivotal question that determines the feasibility of the THz attoclock, requiring definitive experimental clarification.

In this Letter, we demonstrate the 
% {\color{red}
% \sout{first}}
successful implementation of the THz attoclock.
An actively phase-stable Mach–Zehnder (M-Z) interferometer is employed to stabilize and control the relative phase $\varphi$. $\varphi$ is determined via an in situ self-calibration procedure, ensuring the high-fidelity extraction of the THz polarization rotation.
The ellipticity of the $\omega$-$2\omega$ fields is carefully optimized to suppress electron–ion recombination.
Our measurements reveal that the rotation of $\hat{\boldsymbol{E}}_{\mathrm{THz}}$ at fixed $\varphi$ corresponds to a decreasing effective delay with increasing laser intensity, in agreement with time-dependent Schrödinger equation (TDSE) simulations and previous photoelectron attoclock experiments.
These results establish the feasibility of the THz attoclock as a non-vacuum, all-optical, and non-contact approach to probing ultrafast electron dynamics.

In our experimental setup for the THz attoclock (see Supplementary Section \uppercase\expandafter{\romannumeral1} for details), two-color ($\omega$-$2\omega$) co-rotating fields are implemented~\cite{meng2016_zzx_yjm,twofoci,zjj1}, 
%这里可以引用我和赵静静的文章，咱们在太赫兹领域发的文章都要引用。表示咱们深耕这个课题。
%251102，但是之前的双焦点实验用的是双色线偏场。此处是双色圆偏场。场设置不同还能引用吗？----gyj
%可以的，这里引用的原因是告诉审稿人，我们延用了类似的实验装置
ensuring high THz radiation efficiency. The $\omega$ field (800 nm, 60 fs, 1 kHz)
%以前一旦定义了符号，后面就可以一直用这个符号简写。
is derived from a Ti:sapphire laser system. The $2\omega$ field is generated by frequency-doubling the $\omega$ pulse via a $\beta$-barium-borate (BBO) crystal. The two-color beams are split into two arms of the M-Z interferometer via a dichroic mirror, thus enabling independent control of the intensity and polarization states of the two fields. A $\lambda$/2 wave plate and a $\lambda$/4 wave plate are placed in each beam so that the ellipticities of the $\omega$ and $2\omega$ fields can be adjusted to $\epsilon_\omega \approx 0.81$ and $\epsilon _{2\omega} \approx 0.97$, respectively (see Supplementary Section \uppercase\expandafter{\romannumeral2} for the calibration of ellipticity). 
%在补充材料中，为了避免混乱，补充材料可以分章节，如S1, S2...这里可以指明，看补充材料的哪一章节
The major axis of the $\omega$ field is along the $y$ axis. The relative phase $\varphi$ between the $\omega$-$2\omega$ fields is finely adjusted by changing the distance between the BBO and the plasma. To suppress phase jitter between the $\omega$-$2\omega$ beams, we minimize both the optical path length and the number of reflecting mirrors in the M-Z interferometer. Additionally, an active phase-stable system is implemented to stabilize relative phase fluctuations below $\pm 0.02\pi$. The phase fluctuations are presented in Supplementary Section \uppercase\expandafter{\romannumeral4}. 
%这里也可以写一下，索引到补充材料的哪一节。
By maintaining the peak intensity ratio between the $\omega$ field and $2\omega$ field at 3:1, both laser beams are focused collinearly into the Ar gas cell, with coincident focal points, inducing plasma formation and nearly linearly polarized THz wave emission. 
%我把总光强的符号定义在这里了，不知道是否妥当。
$I$ is defined as the sum of the peak intensities of the two-color fields, $I$ = $I_\omega$ + $I_{2\omega}$, where $I_\omega$ and $I_{2\omega}$ represent the peak intensities of the $\omega$ and $2\omega$ fields, respectively. 
The peak electric fields along two orthogonal polarization directions of THz radiation ($E_x$, $E_y$) are measured using the electro-optic sampling (EOS) method with a ZnTe crystal, thereby obtaining the THz polarization direction. The ZnTe crystal is calibrated to have the same response to $E_x$ and $E_y$ (see Supplementary Section \uppercase\expandafter{\romannumeral3} for details).

\begin{figure}[tb]
\centering
\includegraphics[width=8.5cm,height=10.45cm]{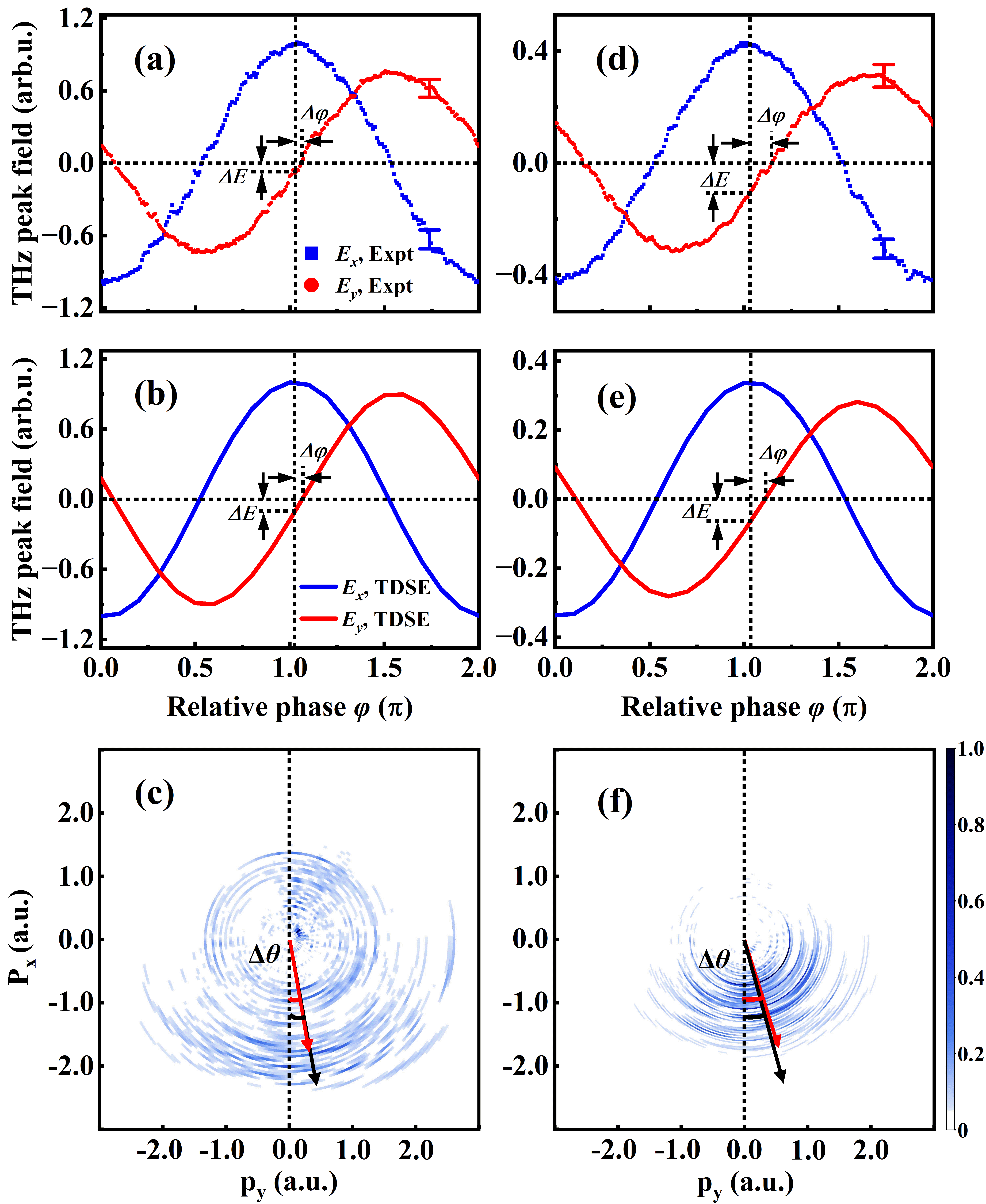}
%我觉得横轴应该标注上\varphi
\caption{\label{fig2} THz yield at different laser intensities $I$. (a,b) Measured and TDSE-calculated THz peak electric fields along the horizontal and vertical polarizations as a function of the relative phase of the two-color fields, $E_x(\varphi)$ (blue) and  $E_y(\varphi)$ (red), at $I = 2.9\times10^{14}~\mathrm{W/cm}^2$. The $\varphi$ axis is calibrated. $\varDelta \varphi$ is defined as the phase difference between the $\varphi$ at the maximum of $E_x$ and that where $E_y = 0$. $\varDelta E$ denotes the absolute value of $E_y$ at the maximum of $E_x$. (c) TDSE-calculated PMD, along with the TDSE-calculated $\hat{\boldsymbol{E}}_{\mathrm{THz}}$ (black arrow) and measured THz polarization direction $\hat{\boldsymbol{E}}_{\mathrm{THz}}$ (red arrow) at $\varphi$ = 0 and $I = 2.9\times10^{14}~\mathrm{W/cm}^2$. The black dotted line in (c) indicates the minor axis direction of the $\omega$ field.
%The black arrow indicates the asymmetric direction of PMD ($\hat{\boldsymbol{P}}_{\mathrm{PMD}}$). 
(d,e) Same as (a,b), but for $I = 1.6\times10^{14}~\mathrm{W/cm}^2$. Error bars in (a) and (d) represent the standard deviation.
(f) Same as (c), but for $I = 1.6\times10^{14}~\mathrm{W/cm}^2$. 
%The theoretical PMD, theoretical $\hat{\boldsymbol{P}}_{\mathrm{PMD}}$ and theoretical $\hat{\boldsymbol{E}}_{\mathrm{THz}}$ are obtained via TDSE simulation. We perform all TDSE simulations on atomic H.
%
%
%
%
%
PMD in (c) and (f) are normalized to its maximum.  
}
\end{figure}
%图e和f等TDSE理论计算的结果。
\iffalse
In the implementation of the THz attoclock, the exact value of the relative phase $\varphi$ between $\omega$-$2\omega$ fields has to be determined. The THz polarization direction $\hat{\boldsymbol{E}}_{\mathrm{THz}}$ via $\omega$-$2\omega$ fields of circular and elliptical polarization rotates as $\varphi$ changes~\cite{linear_phase}. For THz attoclock measurements, the variation of $\hat{\boldsymbol{E}}_{\mathrm{THz}}$ with laser intensity must be compared at a fixed $\varphi$. The exact value of $\varphi$ is experimentally challenging to define precisely.  Adjusting the laser intensity during the experiment may induce slight jitter in $\varphi$. However, even minor variations in $\varphi$ can significantly alter $\hat{\boldsymbol{E}}_{\mathrm{THz}}$, thereby affecting the THz attoclock measurements.
\fi

In implementing the THz attoclock, it is essential to accurately determine the relative phase $\varphi$ between the $\omega$ and 2$\omega$ fields.
%The polarization direction of the THz field, $\hat{\boldsymbol{E}}_{\mathrm{THz}}$, generated by circularly or elliptically polarized $\omega$-$2\omega$ fields, rotates as $\varphi$ varies.
The THz field is generated by circularly or elliptically polarized $\omega$-$2\omega$ fields, and its polarization direction $\hat{\boldsymbol{E}}_{\mathrm{THz}}$ rotates as $\varphi$ varies.
%The THz field generated by circularly or elliptically polarized $\omega$-$2\omega$ fields exhibits a polarization direction $\hat{\boldsymbol{E}}_{\mathrm{THz}}$ that rotates with $\varphi$.
For reliable THz attoclock measurements, the dependence of $\hat{\boldsymbol{E}}_{\mathrm{THz}}$ on laser intensity must therefore be compared at a fixed $\varphi$.
However, precisely defining and maintaining the absolute value of $\varphi$ in an experiment is technically challenging.  Intensity adjustments can introduce phase jitter, and such minor variations in $\varphi$ can lead to pronounced changes in $\hat{\boldsymbol{E}}_{\mathrm{THz}}$, thereby affecting the accuracy of THz attoclock measurements.
Figs.~\ref{fig2}(a) and~\ref{fig2}(d) show the THz peak electric fields in the horizontal and vertical directions as a function of $\varphi$ at different intensities.  
The laser intensity is calibrated by matching the experimentally measured $\Delta\varphi$ with theoretical predictions.
(see Supplementary Section \uppercase\expandafter{\romannumeral7} for details).
Fig.~\ref{fig2}(a) presents $E_x(\varphi)$ and $E_y(\varphi)$ at $I = 2.9\times10^{14}~\mathrm{W/cm}^2$, while Fig.~\ref{fig2}(d) shows the corresponding results at $I = 1.6\times10^{14}~\mathrm{W/cm}^2$. 
Within each panel, the data points of $E_x(\varphi)$ and $E_y(\varphi)$ (blue and red dots) are phase-stable and aligned along the same phase axis $\varphi$, allowing direct comparison.
However, varying the laser intensity can cause slight fluctuations in $\varphi$.
Without phase calibration, the $\varphi$ axes in Figs.~\ref{fig2}(a) and~\ref{fig2}(d) differ, making direct comparison between datasets impossible.
Moreover, without an accurate determination of $\varphi$, the directions of the vector potentials of the $\omega$ and $2\omega$ fields remain undefined, and the resulting angular offset cannot be determined.

We calibrate $\varphi$ according to TDSE simulations based on the well-established Qprop library~\cite{Bauer2020_qprop}. 
The simulations are performed with a maximum orbital angular momentum of up to 100 in the partial-wave expansion to ensure convergence (see Supplementary Section \uppercase\expandafter{\romannumeral6}).
The results show that as the total intensity $I$ of the two-color fields decreases from $4\times10^{14}$ to $1\times10^{14}~\mathrm{W/cm}^2$, the phase at which $E_x$ is maximized lies in the range of $1.02\pi$ to $1.04\pi$ (see Supplementary Section \uppercase\expandafter{\romannumeral5}).
%We calibrate $\varphi$ according to TDSE simulations, which show that as the total intensity $I$ of the two-color fields decreases from $4\times10^{14}$ to $1\times10^{14}$ W/cm$^2$, the phase of maximum $E_x$ lies in the range of $1.02\pi$ to $1.04\pi$ (see Supplementary Section \uppercase\expandafter{\romannumeral5}).
%
We therefore assign the phase of maximum $E_x$ to the mean value of $1.03\pi$. The associated calibration uncertainty is about $0.01\pi$, comparable to the intrinsic phase jitter of the actively phase-stable two-color M–Z interferometer. Using this phase-calibration scheme, we can directly determine the relative phase at the focal region, with phase shifts induced by nonlinear effects inherently taken into account.

\iffalse
After comparing theory and experiment, we define $\varphi$ = $1.03\pi$ as the relative phase where $\mathop{{E}}\nolimits_{{x}} $ reaches its maximum, as indicated by the dotted line in Figs.~\ref{fig2}(a) and~\ref{fig2}(d). The rationale for this calibration is that TDSE simulations show when the total intensity \textit{I} of the two-color fields
decreases from $4\times10^{14}$ W/cm$^2$ to $1\times10^{14}$ W/cm$^2$, $\varphi$ at the maximum $\mathop{{E}}\nolimits_{{x}}$ changes slightly  from $1.02\pi$ to $1.04\pi$ (see Supplementary Sect. \uppercase\expandafter{\romannumeral5} for details). Therefore, we select the mean value of $1.03\pi$ as $\varphi$ corresponding to the maximum value of $\mathop{{E}}\nolimits_{{x}}$. \textit{I} is defined as the sum of the peak intensities of the two-color fields, calculated as \textit{I} = $I_\omega$ + $I_{2\omega}$, where $I_\omega$ and $I_{2\omega}$ represent the peak intensities of the $\omega$ and $2\omega$ fields, respectively. This calibration method introduces an error of  $0.01\pi$, which is close to the error from the phase jitter of the two-color M-Z interferometer.
\fi

In Figs.~\ref{fig2}(a) and~\ref{fig2}(d), $E_x(\varphi)$ and $E_y(\varphi)$ at different intensities $I$ are plotted along the calibrated $\varphi$ axis for direct comparison. The measured $E_x(\varphi)$ and $E_y(\varphi)$ curves exhibit no observable distortion and are in good agreement with the TDSE simulations (Figs.~\ref{fig2}(b), \ref{fig2}(e)), indicating that nonlinear effects do not significantly distort the driving laser field. Both experiments and TDSE simulations show that $E_x$ and $E_y$ increase with $I$. 
%As $I$ decreases, $E_y(\varphi)$ exhibits a relative phase shift with respect to $E_x(\varphi)$, quantified by $\Delta\varphi$ and $\Delta E$ (Fig.~\ref{fig2}). As $I$ decreases, $\varDelta \varphi$ and $\varDelta E$ increase, consistent with the TDSE results.
As $I$ decreases, $E_y(\varphi)$ exhibits a relative phase shift with respect to $E_x(\varphi)$, quantified by $\Delta\varphi$ and $\Delta E$, both of which increase consistently with TDSE results.
Since the THz polarization angle is given by $\theta = \arctan(E_y / E_x)$, variations in $\Delta\varphi$ and $\Delta E$ at different $I$ correspond directly to a rotation of $\hat{\boldsymbol{E}}_{\mathrm{THz}}$. 
%Figs.~\ref{fig2}(c) and~\ref{fig2}(f) show the measured and simulated $\hat{\boldsymbol{E}}_{\mathrm{THz}}$ at different $I$. Theoretical PMD and $\hat{\boldsymbol{E}}_{\mathrm{THz}}$ are obtained using TDSE simulation for the hydrogen atom.
Figs.~\ref{fig2}(c) and~\ref{fig2}(f) show the TDSE-calculated PMD and $\hat{\boldsymbol{E}}_{\mathrm{THz}}$ (black arrow) for the H atoms at different intensities $I$, together with the corresponding measured $\hat{\boldsymbol{E}}_{\mathrm{THz}}$ (red arrow).

In the photocurrent model, both $\Delta\varphi$ and $\Delta E$ vanish, corresponding to a fixed phase difference of $0.5\pi$ between $E_y(\varphi)$ and $E_x(\varphi)$, independent of the laser intensity $I$ (see Supplementary Section \uppercase\expandafter{\romannumeral6}).
%According to this model, $\hat{\boldsymbol{E}}_{\mathrm{THz}}$ remains horizontally at $\varphi = 0$, independent of $I$.
According to this model, $\hat{\boldsymbol{E}}_{\mathrm{THz}}$ remains horizontally at $\varphi = 0$.
%251117,如果写PC模型算得的相位为0处THz偏振方向的话，在补充材料里怎么呈现呢？目前补充材料只有PC与CTMC（考虑库伦势）计算得到的Ex(\varphi)和Ey(\varphi)--------------gyj
In contrast, nonzero and intensity-dependent $\Delta\varphi$ and $\Delta E$ can only be reproduced by TDSE and classical trajectory Monte Carlo (CTMC) simulations~\cite{CTMC1,CTMC2,CTMC3}, where the Coulomb potential is included.
Experimentally, we observe finite and intensity-varying $\Delta\varphi$ and $\Delta E$, providing direct evidence that the THz attoclock successfully probes photoelectron dynamics.
\begin{figure}[tb]
\centering
\includegraphics[width=8.7cm,height=6.25cm]{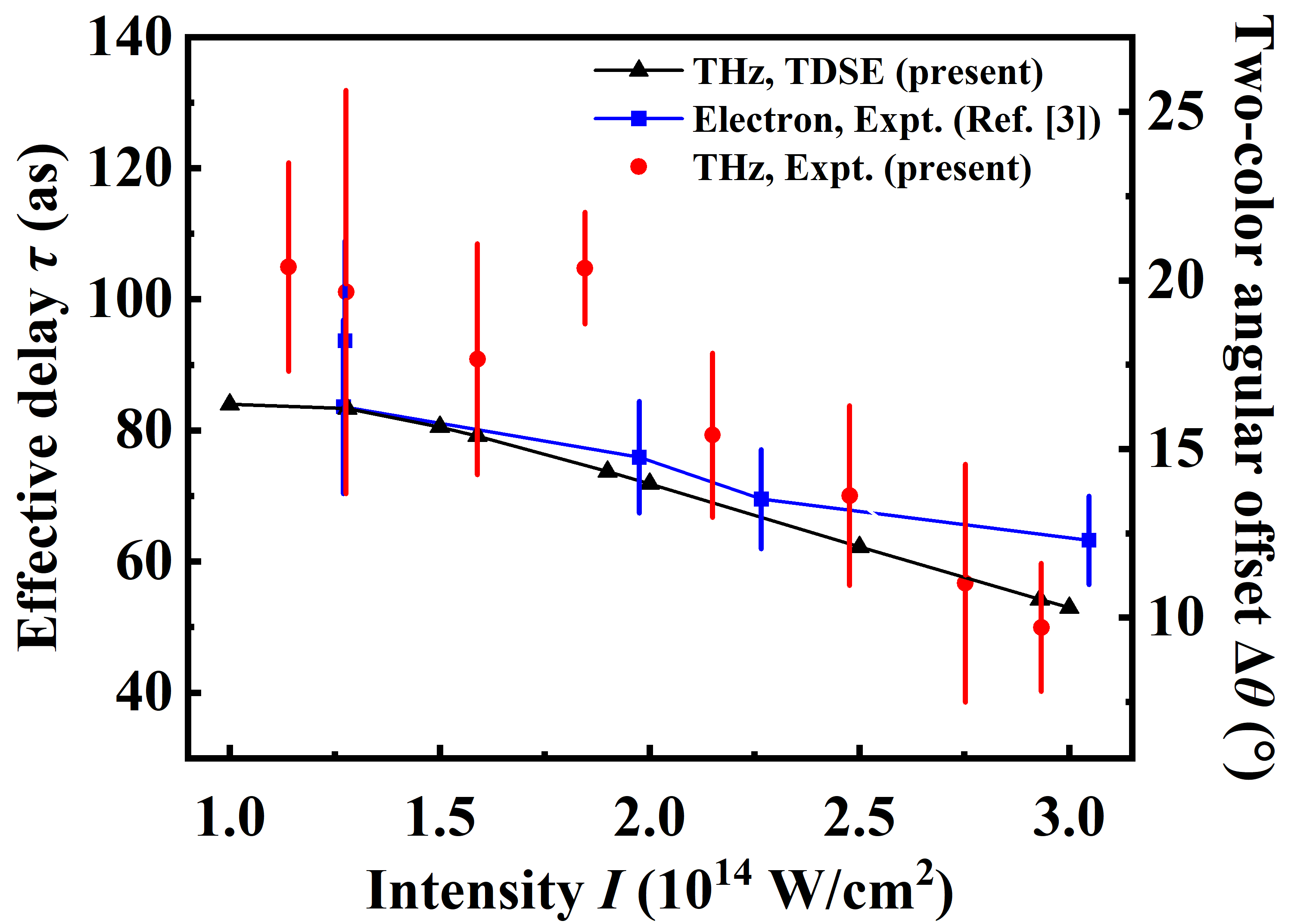}
\caption{\label{fig3}
Comparison of the experimental and TDSE results for the THz attoclock, together with experimental data from the photoelectron attoclock. Red dots show the measured angular offset $\Delta\theta$ of the THz polarization (right $y$-axis) and the corresponding effective delay $\tau$ (left $y$-axis) in Ar as a function of the laser peak intensity $I$ at the $\omega$–$2\omega$ phase $\varphi = 0$, with error bars indicating standard deviations. Black triangles and blue squares denote TDSE simulations of the THz attoclock for H atoms and the effective delay $\tau$ extracted from the single-color photoelectron attoclock in Ar, respectively.
%The right axis shows the $\Delta\theta$ of THz polarization obtained from the experiment and TDSE simulations. Black triangles indicate TDSE simulations for atomic H. Blue squares correspond to the experimental results of photoelectron attoclock, and the effective delay is also extracted from the angular offset.
}
%251110，老师，右轴可以表示实验和TDSE的角偏移。右轴应该怎么画更合适，想问下您的建议-----gyj
%维持现状就可以。
\end{figure}

Fig.~\ref{fig3} presents the angular offset $\Delta\theta$ as a function of the calibrated laser intensity $I$, where $\Delta\theta$ denotes the angle between $\hat{\boldsymbol{E}}_{\mathrm{THz}}$ and the minor axis of the $\omega$ field. The effective delay $\tau$ is extracted using $\Delta\theta \approx 1.44$ $\omega \tau$~\cite{alloptical}. The factor 1.44 is detailed in the Supplementary Section \uppercase\expandafter{\romannumeral8}.
%$\Delta\theta$ arises from the contributions of tunneling ionization delay, nonadiabatic effects, and Coulomb effects during subsequent ionization dynamics, which can be converted into an effective delay $\tau$ by $\Delta\theta \approx 1.44$ $\omega \tau$.
 %Note that $\tau$ does not represent a physical tunneling delay~\cite{alloptical}.
%Because the peak intensity $I$ in focus cannot be measured directly, $I$ is calculated from the pulse energies and the estimated beam waist radii. Using the wavelengths, focal lengths, and pre-focus spot sizes, the waists of the $\omega$ and $2\omega$ fields are estimated as $50~\mu$m and $25~\mu$m, respectively, allowing calibration of $I$ via $I = I_\omega + I_{2\omega}$ and direct comparison with hydrogen TDSE simulations (black triangles).
%251117这里要不写到补充材料里，因为标定光强的方法不够科学-----------gyj
We further compare the THz attoclock with the single-color photoelectron attoclock in Ar~\cite{1attoclock2012_KellerUrsula} (blue squares). Since the latter employs an elliptically polarized field, its PMD angular offset is not directly comparable to the THz polarization angle.
%The PMD angular offset $\Delta\theta'$ is converted to the effective delay $\tau$ as $\Delta\theta' \approx 1.28 \omega \tau$~\cite{alloptical}, enabling a consistent comparison between the two techniques.
%, where $w_0'$ is the frequency of the corresponding single-color field.
%
%251127,老师，这里w是指双色THz阿秒钟里基频场800nm的角频率。而在光电子阿秒钟里，场频率为740nm。并不一样。应该用w'。如下面红色字体所示---------gyj
%
The PMD angular offset $\Delta\theta'$ is converted to the effective delay $\tau$ as $\Delta\theta' \approx 1.28$ $\omega' \tau$~\cite{alloptical}, where $\omega'$ denotes the frequency of the corresponding single-color field, thereby enabling a consistent comparison between the two techniques.
%{\color{red}
%The PMD angular offset $\Delta\theta'$ is converted to the effective delay $\tau$ as $\Delta\theta' \approx 1.28 \omega' \tau$~\cite{alloptical}, enabling a consistent comparison between the two techniques.
%}
The effective delay $\tau$ obtained from the THz attoclock decreases with increasing intensity $I$, consistent with the trend observed in the single-color photoelectron attoclock~\cite{1attoclock2012_KellerUrsula}, which confirms the robustness of the THz attoclock method.

\begin{figure}[tb]
\centering
\includegraphics[width=8.5cm,height=5.96cm]{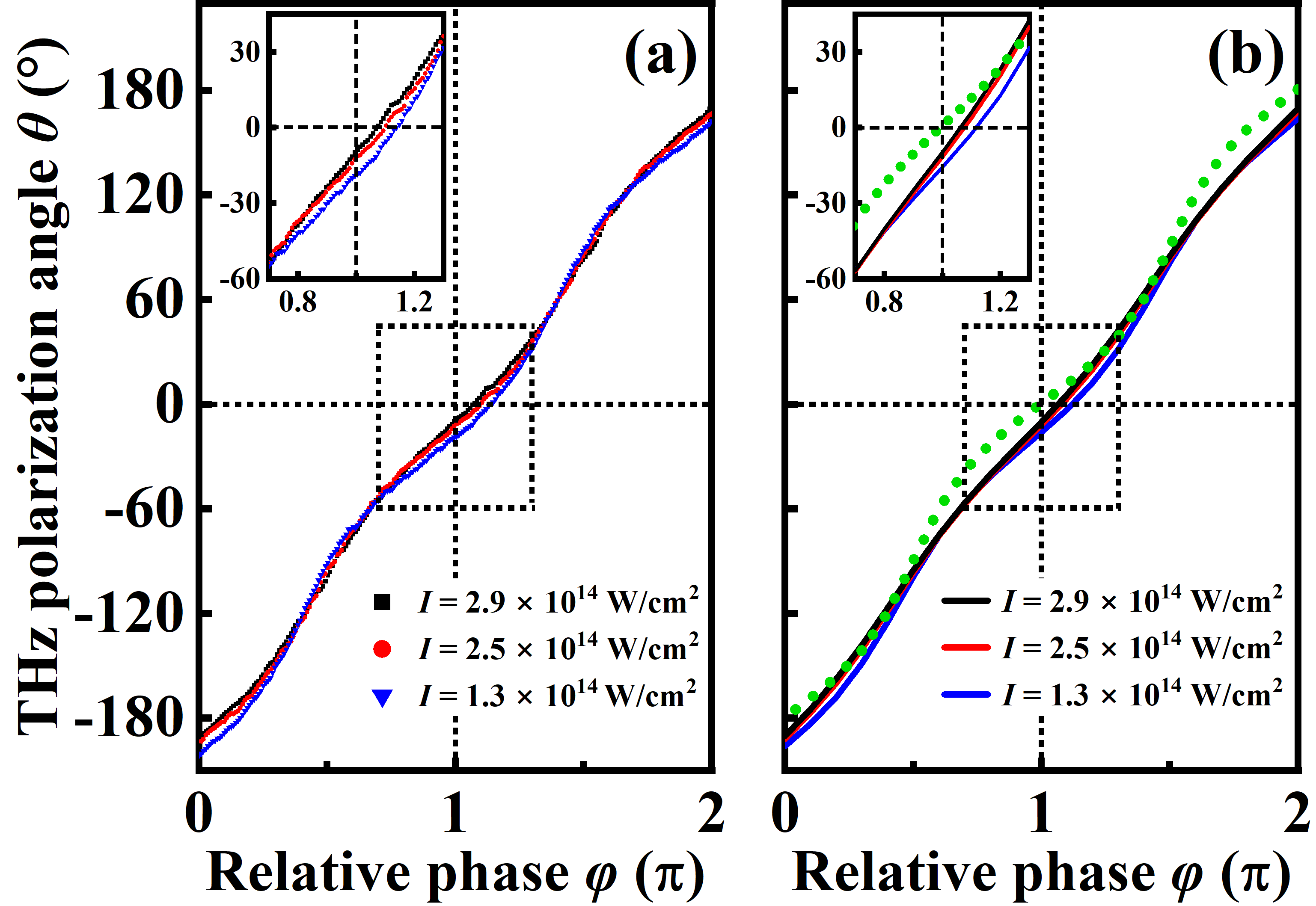}
\caption{\label{fig4} Polarization angle $\theta$ of the THz field $\hat{\boldsymbol{E}}_{\mathrm{THz}}$ as a function of the relative phase $\varphi$ for different laser intensities $I$. (a) Experimental results. (b) TDSE and photocurrent simulations. The solid curves show the TDSE results. The green dotted curve shows the photocurrent prediction at a laser intensity of $1.3\times10^{14}~\mathrm{W/cm}^2$.
}
\end{figure}

While Fig.~\ref{fig3} examines the variation of $\hat{\boldsymbol{E}}_{\mathrm{THz}}$ with laser intensity $I$ at the fixed $\omega$–$2\omega$ phase $\varphi = 0$, Fig.~\ref{fig4} extends the analysis to the full range of $\varphi$ from 0 to $2\pi$.
Figs.~\ref{fig4}(a) and (b) show the measured and TDSE-calculated angle $\theta$ of $\hat{\boldsymbol{E}}_{\mathrm{THz}}$ as a function of $\varphi$ for different $I$. 
%The green dotted line represents the photocurrent model, which neglects the Coulomb potential and predicts $\theta = 0$ at $\varphi = \pi$. 
The photocurrent model (green dotted line), which neglects the Coulomb potential, predicts $\theta=0$ at $\varphi=\pi$.
In contrast, TDSE simulations that include the Coulomb interaction exhibit a zero crossing of the $\varphi$–$\theta$ curves shifted to $\varphi > \pi$. As $I$ increases, the electron escapes more rapidly, reducing the Coulomb influence and causing the $\varphi$–$\theta$ curves to converge toward the prediction of the photocurrent.
%251117，这里光电流的数据目前使用的,1.3e14,要换成较高光强吗----gyj
The close agreement between the experiment and the TDSE simulation validates the robustness of our measurements. Owing to the equivalence between $\hat{\boldsymbol{E}}_{\mathrm{THz}}$ and $\hat{\boldsymbol{P}}_{\mathrm{PMD}}$, the $\varphi$–$\theta$ dependence can be linked to the phase-of-phase curve in PMDs~\cite{2attoclock2021_Liuyunquan_POP}, offering a pathway to reconstruct tunneling dynamics in future work.

In summary, we demonstrate the feasibility of the THz attoclock for probing strong-field electron dynamics.
By measuring the orthogonal THz yields $E_x(\varphi)$ and $E_y(\varphi)$, the relative phase $\varphi$ is self-calibrated, enabling direct comparison of the THz polarization direction $\hat{\boldsymbol{E}}_{\mathrm{THz}}$ across different laser intensities $I$. The effective delay extracted $\tau$ decreases with increasing $I$, consistent with TDSE simulations and photoelectron attoclock measurements. The measured dependence $\varphi$–$\theta$ also converges toward the classical prediction of the photocurrent at higher intensities, further confirming the robustness of the method.
%{\color{red}251127，“收敛于经典光电子预测 ”，觉得不太准确，CTMC也是经典光电子预测，但是实际上收敛于PC模型的结果。}
These results establish the all-optical THz attoclock as a vacuum-free, contactless probe of tunneling electron motion, offering a promising route toward studying laser-driven electron dynamics in condensed-matter systems.

\begin{acknowledgments}
This work is supported by the National Key Research and Development Program of China (No. 2022YFA1604302), the National Natural Science Foundation of China (Nos. 12334011, 12374262, 12174284 and 12541505) and the Natural Science Foundation of Henan (Grant No. 252300421304). We also acknowledge support from the
Shanghai-XFEL beamline project (SBP) (Grant No.
31011505505885920161A2101001) and the computational resource of the HPC platform in ShanghaiTech University.
\end{acknowledgments}

\nocite{*}

\bibliography{apssamp}% Produces the bibliography via BibTeX.

\end{document}